\documentclass[12pt]{article}
\usepackage{epsfig}

\parskip=\medskipamount	
\renewcommand{\thesection}{\arabic{section}}

\catcode`@=11

\@addtoreset{equation}{section}
\@addtoreset{equation}{subsection}
\def\theequation{\ifnum\value{section}=0 \arabic{equation}\ignorespaces
\else \ifnum\value{section}=-1 A.\arabic{equation}\ignorespaces
\else \ifnum\value{subsection}=0 \thesection.\arabic{equation}\ignorespaces
\else \thesection.\arabic{subsection}.\arabic{equation}\ignorespaces
                             \fi
                        \fi
                   \fi}

{\catcode`\'=\active \def'{{}^\bgroup\prim@s}}

\catcode`@=12

\newcommand{\bq}{\begin{equation}}
\newcommand{\be}{\begin{equation}} 
\newcommand{\fq}{\end{equation}}
\newcommand{\ee}{\end{equation}}
\newcommand{\bqr}{\begin{eqnarray}}
\newcommand{\beqs}{\begin{eqnarray}} 
\newcommand{\fqr}{\end{eqnarray}}
\newcommand{\eeqs}{\end{eqnarray}}

\newcommand{\rf}[1]{(\ref{#1})}

\def\bop#1{\setbox0=\hbox{$#1M$}\mkern1.5mu
	\vbox{\hrule height0pt depth.04\ht0
	\hbox{\vrule width.04\ht0 height.9\ht0 \kern.9\ht0
	\vrule width.04\ht0}\hrule height.04\ht0}\mkern1.5mu}

\begin{document}
\thispagestyle{empty} 

\begin{flushright} 
\begin{tabular}{l} 
ANL-HEP-PR-01-012 \\
hep-th/0103056 \\ 
\end{tabular} 
\end{flushright}  

\vskip .3in 
\begin{center} 

{\Large\bf Open string decoupling and tachyon condensation} 

\vskip .3in 

{\bf Gordon Chalmers} 
\\[5mm] 
{\em Argonne National Laboratory \\ 
High Energy Physics Division \\ 
9700 South Cass Avenue \\ 
Argonne, IL  60439-4815 } \\ 

{e-mail: chalmers@pcl9.hep.anl.gov}  

\vskip .5in minus .2in 

{\bf Abstract}  
\end{center}  

The amplitudes in perturbative open string theory are examined as functions of 
the tachyon condensate parameter.  The boundary state formalism demonstrates 
the decoupling of the open string modes at the non-perturbative minima of the 
tachyon potential via a degeneration of open world-sheets and identifies 
an independence of the coupling constants $g_s$ and $g_{YM}$ at general values 
of the tachyon condensate.  The closed string sector is generated at the quantum 
level; it is 
also generated at the classical level perturbatively through the condensation of 
propagating open string modes on the D-brane degrees of freedom.  

\setcounter{page}{0} 
\newpage 
\setcounter{footnote}{0} 

\section{Introduction} 

Open string field theory contains closed strings in its quantum spectrum.  
It is conjectured that open string states decouple at a non-perturbative 
minima of the tachyon condensate \cite{Sen:1999mh} (further analyzed in 
\cite{Sen:1999xm, Sen:2000kd,Bergman:2000xf}).  Approximations of the tachyon 
potential in string field theory are found 
in,\footnote{Investigations of the tachyon potential in the context of dual 
models are found in \cite{Bardakci:1975ux}} 
\cite{Kostelecky:1990nt,Kostelecky:1996qk,Berkovits:2000hf,DeSmet:2000dp, 
Sen:2000nx, Berkovits:2000zj, Moeller:2000xv,Taylor:2000ek,Kutasov:2000qp, 
Gerasimov:2000zp}, for example.  Similar conjectures for the superstring are 
covered in \cite{Sen:1998ii,Sen:1998sm,Sen:1998ex}.  In this letter,  
we examine scattering amplitudes in the mixed open/closed theory within the 
boundary state formalism developed in a number of works \cite{Witten:1993cr, 
Witten:1992qy, Harvey:2000na,Kutasov:2000aq,Kraus:2000nj}; in this formalism we 
may explicitly demonstrate that scattering amplitudes containing open string modes 
decouple, or, equivalently, that Riemann surfaces with non-vanishing boundaries 
describing the worldsheet are exponentially suppressed with respect to the tachyon 
condensate; and we answer a question regarding how the closed string modes arise at 
the classical level.  We also determine independence of the bosonic coupling 
constants in open string theory, $g_s$ and $g_{\rm YM}$, and model them via bulk and 
boundary worldsheet expectation values.  This independence of the string couplings 
generates broken conformal invariance that is found in off-shell open string 
field theory.  

The open/closed string duality \cite{Dolen:1968jr} generates closed strings in 
the quantum amplitudes of the open string theory (and also generates precise means 
to map non-abelian gauge theory amplitudes to gravitational amplitudes 
\cite{Kawai:1986xq}).  It 
is of interest to generate alternate means for finding closed string excitations 
including gravitation from a classical open string model.  The role of the coupling 
constant relation $g_s=g^2_{\rm YM}$ in on-shell open string theory and the 
holomorphic/anti-holomorphic factorization of scattering amplitudes permits a 
direct comparison between unintegrated graviton amplitudes with gauge boson 
amplitudes.  In the context of the BSFT (boundary string field theory) formalism, 
however, the coupling constant relation may be generalized: a mechanism for 
generating a closed string sector with non-interacting open strings is to take 
$g_{\rm YM}\rightarrow 0$.  The boundary worldsheet interactions and the relation 
to the open string coupling constant breaks conformal invariance of the world-sheet 
theory, but may be related to a field redefinition in the target space-time.  This 
limit in the target spacetime is akin to examining 
\bqr  
{\cal L} = \int d^{26}x \sqrt{g} \bigl( {1\over \kappa^2} R + 
  {1\over g_{\rm YM}^2} {\rm Tr} F^2 \bigr) \ ,  
\label{mixedtarget} 
\fqr 
the limit of fixed $\kappa$ and varying $g_{\rm YM}^2$ at fixed $\alpha'$.  In 
the target spacetime theory, it is obvious that general values of the coupling 
constant generate consistent theories.  This will be given a concrete description in 
terms of the open string theory in a non-trivial tachyon background in this 
letter.  A nullified gauge coupling constant decouples open string modes.

Geometrically, there are two scalar deformations on the open string worldsheet that 
generate gauge coupling constants: 
\bqr 
\int_{M_{g,h}} d^2z \sqrt{g} R \phi + 
 \int_{\partial M_{g,h}} dz ~e  \tilde\phi \ , 
\fqr 
with $R$ the bulk worldsheet curvature, the integral  
giving a topological invariant on genus g, $\int d^2z \sqrt{g} R = 
2-2g-h$.  The latter picks up a factor under non-area preserving diffeomorphisms, 
and the scaling may be absorbed in general under a scale change of the coupling 
$\tilde\phi$.  Modulo the 
worldsheet renormalizations, these expectation values couple as closed and open 
string dilaton and the latter models the off-shell structure of perturbation 
theory in the open string theory via the boundary state field formalism.  The 
path integral quantization is direct and may be performed explicitly in 
perturbation theory.  The open string field theory with an off-shell tachyon 
is modeled by the critical bosonic theory with the above deformations.  The 
introduction of the open string boundary term generates both off-shell open 
string theory in addition to the equivalent interpretation in which $g_s\neq 
g_{ym}^2$.  The latter may be interpreted as a breaking of the conformal 
invariance of the string away from the point in which $g_{ym}=0$ ($\tilde 
\phi = -\infty$) and $g_s\neq 0$ in which the closed sector remains as the 
consistent closed bosonic theory.  This holds in perturbation theory order 
by order in the worldsheet expansion.  

This work is organized as follows.  In Section 2 we examine the scattering amplitudes 
in the BSFT formalism and probe them as a function of the vacuum values $\phi$ and 
$\tilde\phi$.  In section 3, we demonstrate that closed strings arise at the 
classical level in the non-perturbative regime of the tachyon vacuum.  In section 
4 we conclude with discussion.  

\section{Amplitudes in the BSFT of the string field} 

The worldsheet bosonic action of the bosonic string theory is described by 
\bqr  
S_0 = \int_M d^2z \sqrt{g} g^{\alpha\beta} \partial_\alpha X^\mu \partial_\beta 
X^\nu G_{\mu\nu} +  \int_M d^2z \sqrt{g} R \phi \ . 
\fqr 
The deformations that breaks worldsheet conformal invariance, 
\bqr  
S[u_j,a]= \int_{\partial M} dz e \bigl(-a+\sum_{j=1}^{26} u_\mu X_\mu^2 \bigr)  \ ,   
\fqr 
describe the solitonic configuration \cite{Sen:2000tg}.  The boundary conditions 
following from the above $S+S[u_j,a]$ are 
\bqr 
n^\alpha \partial_\alpha X_\mu + u_\mu X_\mu = 0 .  
\fqr 
On a surface with boundary the former term arises from the boundary term, 
\bqr  
\int d^2z \sqrt{g} \partial^\alpha \bigl( g^{\alpha\beta} \partial_\beta \chi 
 \bigr) = \int dz e ~a \ , 
\fqr 
and represents the total derivative and boundary fluctuations on the open 
string worldsheet.  

The tachyon potential in the open bosonic string theory has the approximate 
form \cite{Minahan:2000ff} 
\bqr  
V(T)=\exp{\bigl( -a\bigr)} (1+a) \ . 
\fqr 
A target spacetime redefinition, $\exp(-a)=\phi^2$, changes the potential into  
\bqr  
V(\phi) = \phi^2(1-2\ln \phi) \ .  
\fqr 
Within the boundary state formalism, the tachyon condenses at the point $a\rightarrow\infty$ 
and $u_j\rightarrow 0$ (destabilizing the solitonic configuration), modeled by 
exponentiating zero momentum insertions.  

The non-vanishing values of $a$ and $u_j$ break the conformal invariance on the 
boundaries of the Riemann surfaces.  Greens functions are independent of the 
parameter $a$.  However, as the variation of the boundary cosmological constant 
term is independent of $X_\mu$.  The conformal mappings change the functional 
form of the Greens functions 
as a function of $u_j$.  On the disc with flat metric $ds^2=dz^2+d{\bar z}^2$ the 
two-point function with boundary conditions is 
\bqr  
G(z_i,z_j;u) = -\ln\vert z_i-z_j\vert^2 - \ln\vert z_i{\bar z_j}+1\vert^2 + 
{2\over u} 
\fqr 
\bqr 
-2u \sum_{k=1}^\infty {1\over k(k+u)} ~ \Bigl[ \left(z_i {\bar z}_j\right)^k + 
 \left({\bar z}_i z_j\right)^k \Bigr] 
\fqr 
and on the stereographically projected sphere the Greens function, following from a 
coordinate transformation   
\bqr 
z \rightarrow {1-z\over 1+z} 
\fqr 
takes on the form, 
\bqr  
G(z_i,z_j;u) = -\ln\vert z_i-z_j\vert^2-\ln\vert z_i+z_j\vert^2 + {2\over u} 
\fqr 
\bqr 
- 2u 
\quad \sum_{k=1}^\infty {1\over k(k+u)} \left[ \left([1-z][1-{\bar w}]\over 
[1+z][1+{\bar w}] \right)^k  + \left([1-{\bar z}][1-w]\over [1+{\bar z}][1+w] 
\right)^k  \right]  \ . 
\label{twopointfunction}
\fqr 
The limit in which $u_j=0$ generates conformally invariant Greens functions 
(after subtracting the zero mode) as in the critical string.  The limit of 
these Greens functions on the boundary is 
\bqr  
G(\theta_1,\theta_2;u) = 2 \sum_{k\in Z} {1\over \vert k\vert + u} e^{i(\theta_1 
-\theta_2)}  \ , 
\fqr 
together with the first derivative, 
\bqr  
\partial G(\theta_1,\theta_2;u) = 2i \sum_{k\in Z} \left[ k\over \vert k\vert+ 
u\right]  e^{i(\theta_1-\theta_2)} \ .    
\fqr 
The OPE associated with the two-point function is straightforward to carry out and 
generates an identical structure to the undeformed case.   Similar Greens functions 
on multi-genus Riemann surfaces are obtained from the prime form $E(z_i,z_j)$.  To 
explore the tachyon condensation limit, however, we require $u_j=0$ and the 
multi-genus Greens functions are conformally invariant in this limit explicitly 
(they do not depend on the parameter $a$).  

The scattering amplitudes are,  
\bqr  
A(k_j,\epsilon_j) = \int {[dg][dX]\over N} ~e^{-(S_0+S[u_j,a])/\alpha'} ~ 
  \prod_{m=1}^n V_{k_m} \ ,  
\fqr 
modulo a renormalization of the boundary coupling compensating for the conformal 
invariance being broken at the boundary.  The local OPE generates poles identical 
to the free theory, and the mass levels do not shift as a function of $u_j$: they 
are trivially independent of the deformation parameter $a$, because the Greens 
functions are independent of this parameter.  The term $S[0,a]$ appears to break 
conformal invariance; the choice of the gauge fixing of the diffeomorphisms generates 
different exponential factors associated with this term in the world-sheet theory.  
In the target spacetime this ambiguity from 
\bqr 
\exp{\bigl( -a \int dz ~e + S_0\bigr)} \ ,  
\fqr  
is related to the ambiguity in defining an off-shell extension of string theory.  
However, for any choice of metric on the world-sheet a field redefinition of $a$ 
may be performed that removes the ambiguity.  In the target space-time theory this 
is a relabeling of coordinates that parameterizes the moduli space of tachyon 
vacua.  In the target space-time theory, the interaction appears conformal 
invariant.  

Thus the interaction generates on a worldsheet, 
\bqr 
\int dg \exp{\Bigl(-a\int dz ~e\Bigr)} = \exp{\Bigl( -a_{\hat g} (1+h)\Bigr)} \ ,  
\fqr 
where $\hat g$ denotes the conformal class, and $a_{\hat g}$ is a redefined tachyon 
expectation value. 

As an example, consider the four-point function $A_4(k_i,\epsilon_i)$ evaluated on 
the sphere with non-vanishing $u_j$ parameters.  We take the limit in which $u_j=0$; 
the conformal breaking only enters into the conformal prefactor associated with the 
open string coupling.  The Koba-Nielson representation of the amplitude is 
\bqr  
A= e^{-a l} \int \prod_{i,j=1}^4 \exp{\Bigl(-\alpha k_i\cdot k_j G_{ij} + 
 \epsilon_{[i}\cdot k_{j]} {\dot G}_{ij} + \epsilon_i\cdot \epsilon_j 
 {\dot{\dot G}}_{ij}\Bigr)} \vert_{\rm multi-linear}  
\fqr 
and as an expansion in the polarizations it generates the tree-level amplitude 
process.  The cosmological constant term in the exponential from $S[0,a]$, $al$, 
generates the exponential suppression at the boundary and thus the OPE is 
independent of $a$.  

The gauge fixed $n$-point tachyon amplitude is similarly, 
\bqr 
A_4 = e^{-a\beta} ~(z_{n-2}-z_{n-1})(z_{n-1}-z_n)(z_{n-1}-z_n) \nonumber 
\fqr 
\bqr 
\int\prod dz_j 
 \prod_{i\neq j} \exp{\Bigl( -G_{ij} s_{ij}\Bigr)} \vert_{z_1=0,z_2=1,z_n=\infty} 
\fqr 
with $s_{ij}=(k_i+k_j)^2$.  Mixed scattering 
between the gauge bosons and tachyon modes are found by replacing the $j$th 
polarization vector in the above with zero and substituting in the mass shell 
condition of the tachyon, $k_j^2=m^2$.  

The expansion of the perturbative series in the genus expansion is 
\bqr  
A_n(k_i,\epsilon_i;a) = \sum e^{-p\phi} e^{-a_{\hat g}} A_n^{(m)}(\epsilon,k_i) \ , 
\fqr 
with $\phi$ the closed dilatonic factor, and $A_n^{(m)}$ the $n$-point genus $m$ 
sub-amplitude.  Consistency with unitarity requires the relative normalization between 
the open couplings of $e^{-a_{\hat g}}$ at different orders in the expansion.  

Next, we comment on the behavior of the poles and residues as a function of the 
tachyon vacuum at non-vanishing values of $u$.  The Greens function in 
\rf{twopointfunction} has the expansion at small values of $u$ 
\bqr  
e^{\alpha' G(z_1,z_2)} = \vert z_1-z_2\vert^{2\alpha'} 
 \vert z_1+{\bar z}_2\vert^{2\alpha'} e^{2/u} \nonumber 
\fqr 
\bqr  
\times \sum_{m=0}^\infty (-2u)^m \Bigl[ \sum_{k=1}^\infty {1\over k(k+u)} 
  \Bigl( {(1-z) (1-{\bar w})\over (1+z)(1+{\bar w})} \Bigr)^k 
 + \Bigl( {(1-{\bar z})(1-w) \over (1+{\bar z})(1+w) }\Bigr)^k \Bigr]^m  \,
\fqr 
and the poles do not shift as a function of $u$.  However, the residues of the poles 
have a non-trivial $u$ structure, and the expansion of a pole changes as 
\bqr  
\sum_{k=1}^\infty {\alpha(k,u;s)\over s-m_k} \ ,  
\fqr 
with $\alpha(k,u;s)=b_n+\alpha n + g(u)+n h(u)$.  This signals that the tachyon 
medium has a dielectric effect on the angular momentum associated with the pole.  

\section{Closed strings via confinement on the brane}

Nonperturbative mechanisms have been analyzed inducing confinement of the open 
strings, see for example \cite{Bergman:2000xf,Gerasimov:2001ga}.  The presence 
of the Dirichelet boundary conditions reflect a soliton in the spacetime, and 
we present two microcopic mechanisms of the dynamics of the open strings into 
closed counterparts.  The first mechanism reflects the Hilbert space.  The 
second represents the forces acting on the endpoints of the open strings and 
not the closed strings themselves; however, the force in the latter mechanism 
does indicate a confinement into closed strings.
  
One mechanism for the generation of closed strings at the classical level is found 
through the limit $a\rightarrow \infty$.  The open strings propagating on the soliton 
parameterized by the Dirichelet boundary conditions span a worldsheet with boundary, 
and as such must degenerate in the limit of $a\rightarrow\infty$ (since the coupling 
is $e^{-a l_{\rm bdy}}$)\footnote{This is also examined in the context of ``hole 
cutting operators'' in \cite{Gerasimov:2001ga}}.  This limit provides a dynamical 
means for the doubling of the Hilbert space at the classical level.  

Furthermore, at non-vanishing values of $u$, the force between two points on the 
soliton may be computed from the tree-level Greens functions $G(z_1,z_2;a,u)$.  At 
values $u\rightarrow 0$, the Greens function creating the force between two test 
particles on the brane is 
\bqr 
G(z_1,z_2;u) = -\ln\vert z_1-z_2\vert^2 - \ln\vert z_1{\bar z}_2 + 1\vert^2 + 
{2\over u} 
+ {\cal O}(u)  \ . 
\fqr  
As $u\rightarrow 0$ a potential from the zero mode binds the two 
points on the soliton, closing the open string degrees of freedom on the soliton 
as $u\rightarrow 0$ and as $a\rightarrow \infty$.  This confining mechanism (linear 
in $1/u$) dynamically generates the closed string degrees of freedom at the 
classical level, as the tachyon condenses and the brane evaporates.  

\section{Conclusions}

The perturbative amplitudes in the boundary state formalism are  examined together 
with the pole structure of the amplitudes and the functional dependence 
of $a$ and $u_j$.  In the limit in which $a\rightarrow\infty$, the amplitudes 
associated with bounded Riemann surfaces (i.e. open string worldsheets) are 
exponentially suppressed as a function of the tachyon expectation value.  In the 
limit of the tachyon mimima, only the closed string sector remains, in agreement 
with the Sen conjecture.  The mechanism for the decoupling of the open strings is 
perturbative in the boundary state formalism, and generalizes to off-shell the 
open-closed duality.  An interpretation of the parameter $a$ is a decoupling of 
the two couplings $g_s$ and $g_{\rm YM}^2$ in the string, which leads to a breaking 
of conformality on the world-sheet.  The ambiguity associated with the world-sheet 
non-conformal invariant term, $-a\int dz e$, along the boundaries of the Riemann 
surfaces may be absorbed by a field redefinition of the tachyon expectation value 
labeling the vacua.  This procedure holds order by order in perturbation theory 
and multi-genus results may be performed in a similar fashion.   

The tachyon condensation in this framework indicates that the closed string 
excitations are perturbative.  The closed string modes arise as bound states of 
the open string states propagating on the brane as the tachyon condenses.  The 
decoupling of the open worldsheet surfaces generates a dynamical means for doubling 
the Hilbert space, as is required for consistency with quantum open/closed duality.

\vskip .3in 
\section*{Acknowledgements} 

The work of GC is supported in part by the US Department of Energy, Division of 
High Energy Physics, contract W-31-109-ENG-38.  I thank Per Kraus and Finn Larsen 
for discussions.

\end{document}